\documentclass[superscriptaddress,prb,amsmath,amssymb,twocolumn,aps]{revtex4-1}
\usepackage{graphicx}
\usepackage{booktabs}
\usepackage{xcolor}

\usepackage{hyperref}
\hypersetup{colorlinks=true,citecolor=blue,linkcolor=blue,urlcolor=blue}
\begin{document}

\title{The doping evolution of the charge density wave and charge density fluctuations in La$_{2-x}$Sr$_x$CuO$_4$}

\author{Charles C. Tam}
\email{c.tam@bristol.ac.uk}
\affiliation{H.\;H. Wills Physics Laboratory, University of Bristol, Bristol BS8 1TL, United Kingdom.}
\affiliation{Diamond Light Source, Harwell Campus, Didcot OX11 0DE, United Kingdom.}
\author{Mengze Zhu}
\altaffiliation[Current address: ]
{Laboratory for Solid State Physics, ETH Zürich, 8093 Zürich, Switzerland.}
\affiliation{H.\;H. Wills Physics Laboratory, University of Bristol, Bristol BS8 1TL, United Kingdom.}
\author{Maud C. Barthélemy}
\affiliation{H.\;H. Wills Physics Laboratory, University of Bristol, Bristol BS8 1TL, United Kingdom.}
\author{Lauren J. Cane}
\affiliation{H.\;H. Wills Physics Laboratory, University of Bristol, Bristol BS8 1TL, United Kingdom.}
\author{Oliver J. Lipscombe}
\affiliation{H.\;H. Wills Physics Laboratory, University of Bristol, Bristol BS8 1TL, United Kingdom.}
\author{Stefano Agrestini}
\affiliation{Diamond Light Source, Harwell Campus, Didcot OX11 0DE, United Kingdom.}
\author{Jaewon Choi}
\affiliation{Diamond Light Source, Harwell Campus, Didcot OX11 0DE, United Kingdom.}
\author{Mirian Garcia-Fernandez}
\affiliation{Diamond Light Source, Harwell Campus, Didcot OX11 0DE, United Kingdom.}
\author{Ke-Jin Zhou}
\altaffiliation[Current address: ]
{National Synchrotron Radiation Laboratory and School of Nuclear Science and Technology, University of Science and Technology of China, Hefei 230026, China.}
\affiliation{Diamond Light Source, Harwell Campus, Didcot OX11 0DE, United Kingdom.}
\author{Stephen M. Hayden}
\email{s.hayden@bristol.ac.uk}
\affiliation{H.\;H. Wills Physics Laboratory, University of Bristol, Bristol BS8 1TL, United Kingdom.}

\begin{abstract}

Cuprate superconductors show various collective charge correlations that are intimately connected with their electronic properties. In particular, charge order in the form of an incommensurate charge density wave (CDW) order with an in-plane wavevector $\delta_{\text{CDW}} \approx $ 0.23--0.35~r.l.u. appears to be universally present. In addition to CDW, dynamic charge density fluctuations (CDF) are also present with wavevectors comparable to $\delta_{\text{CDW}}$.  CDFs are present up to $\sim300\;$K and have relatively short correlation lengths of $\xi \sim 20$\;\AA.  Here we use Cu-$L_3$ and O-$K$ resonant inelastic X-ray scattering (RIXS) to study the doping dependence of CDW and CDFs in La$_{2-x}$Sr$_x$CuO$_4$. We fit our data with (quasi)elastic peaks resulting from the CDW and up to four inelastic modes associated with oxygen phonons that can be strongly coupled to the CDFs.  Our analysis allows us to separate the charge correlations into three components: the CDW with wavevector $\delta_{4a-\text{CDW}} \approx 0.24$ and two CDF components with $\delta_{4a-\text{CDF}} \approx 0.24$ and $\delta_{3a-\text{CDF}} \approx 0.30$.  We find that for $T \approx T_c$ the CDW coexists with the CDFs for dopings near $x=p \sim 1/8$.  The $4a$-CDW disappears beyond $x=0.16$ and the $4a$-CDF beyond $x=0.19$, leaving only a weak $3a$-CDF at the highest doping studied, $x=0.22$.  Our data suggest that low-energy charge fluctuations exist up to doping $x=0.19=p^{\star}$, where the pseudogap disappears, however, we find no evidence that they are associated with a quantum critical point.   
\end{abstract}

\maketitle

\section{Introduction}
\label{sec:intro}
Understanding the normal state from which superconductivity emerges is essential to constructing theories of high-$T_c$ superconductivity in the cuprates. The normal state of the hole-doping versus temperature $p$-$T$ phase diagram of the cuprates contains a number of intertwined phases and behaviors~\cite{Keimer_N_2015}. These include the pseudogap, strange-metal behavior, and spin and charge density wave order (SDW, CDW) and their associated fluctuations. In this paper we focus on CDW order and charge density fluctuations (CDF) \cite{Hayden2024_HT}.  Our aim is to determine their evolution with doping, so that they may be correlated with other physical properties.  

Conventional CDWs are periodic modulations of the valence charge density which have a corresponding atomic displacement\cite{Monceau2012_Mon}. In conventional continuous phase transitions, there is an order parameter whose correlation length $\xi$ diverges on cooling through the transition temperature. The charge correlations in cuprates do not follow this behavior\cite{Hayden2024_HT}.    
At high temperatures, $T \sim $150--300\;K (depending on the system), shortrange correlations with $\xi \sim 15$~\AA\ and in-plane wavevectors $\mathbf{Q}=(\delta,0)$ and $(0,\delta)$ with $\delta \sim $0.23--0.35~r.l.u. are observed \cite{Hayden2024_HT,Ghiringhelli_S_2012, Tabis_PRB_2017, Lee_NP_2020}. We will loosely call these correlations ``charge density fluctuations'' (CDF). On cooling a second component which we call ``charge density wave order'' (CDW) develops below a characteristic temperature $T_{\text{CDW}}$. The correlation length and the intensity $I_{\text{CDW}}$ of the CDW order grows on cooling with $\xi_{\text{CDW}}$ reaching approximately 70~\AA\ in YBa$_2$Cu$_3$O$_{6+x}$ (YBCO) \cite{Ghiringhelli_S_2012, Chang_NP_2012}, 35~\AA\ in La$_{2-x}$Sr$_x$CuO$_4$ (LSCO)~\cite{Croft_PRB_2014, Thampy_PRB_2014} and 160~\AA\ in Tl$_2$Ba$_2$CuO$_{6+x}$ (Tl2201)~\cite{Tam_NC_2022}. Below $T_c$, both $\xi_{\text{CDW}}$ and $I_{\text{CDW}}$ can be reduced. This behavior is interpreted as evidence of competition between CDW and superconductivity.
In LSCO, the CDW peak is relatively narrow at low temperature $T=T_c \approx31\;$K with $\xi\sim 35$~\AA\ and becomes broader with $\xi\sim15$~\AA\ for  $T \approx 70\;$K \cite{Croft_PRB_2014,Thampy_PRB_2014}. 

Resonant inelastic X-ray scattering (RIXS) can be used to investigate collective charge correlations. RIXS is more sensitive to the valence charge density than non-resonant X-ray methods. The state-of-the-art energy resolution of soft X-ray RIXS spectrometers is approaching 10\;meV at the time of writing \cite{Huang_PRX_2021}, meaning quasi-elastic or elastic scattering components can be resolved from inelastic scattering. Previous RIXS studies on cuprates have identified separate CDW and CDF components in YBCO\cite{Arpaia_S_2019}, HgBa$_2$CuO$_{4+\delta}$ (Hg124)\cite{Yu_PRX_2020} and LSCO\cite{Huang_PRX_2021} with the CDW contribution being sharper in reciprocal space than the CDF.    

\begin{figure*}[htb!]
\includegraphics[]{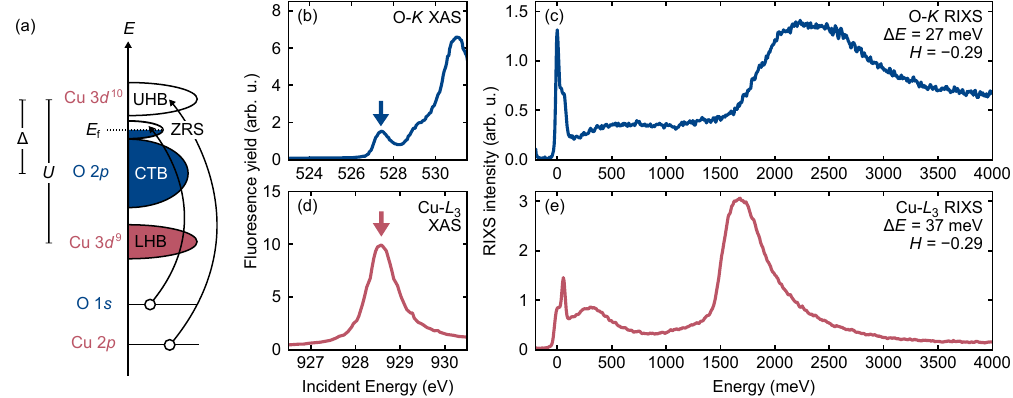}
\caption{\textbf{RIXS at two atomic sites.} (a) Schematic energy level diagram of superconducting cuprates in the Zaanen-Allen-Swartzky scheme \cite{Zaanen_PRL_1985}, and the excitations that Cu-$L_3$ and O-$K$ RIXS creates within the scheme. UHB means upper-Hubbard band, ZRS means Zhang-Rice singlet, and LHB means lower-Hubbard band. $\Delta$ is the charge-transfer energy and $U$ is the Coulomb energy. (b), (d) Cu-$L_3$ and O-$K$ XAS spectra obtained in fluorescence yield mode. The arrows mark the incident energy for the RIXS experiments. Small discrepancies between absolute energies in the XAS spectra and published values are due to a small miscalibration in the beamline monochromator. (c), (e) RIXS spectra taken at $H=-0.29$ at Cu-$L_3$ and O-$K$ edges respectively. All measurements are taken on the same $x=0.19$ LSCO sample.}
\label{fig:overview} 
\end{figure*}

In order to understand the physical properties and phase diagram of the cuprates, it is important to know how the various components of the charge correlations described above evolve with doping. LSCO is well suited for this type of study as it can be doped through the phase diagram. Thus, in this article we investigate the nature of the CDW order and CDFs in La$_{2-x}$Sr$_x$CuO$_4$ as a function of doping.  There have been a number of studies\cite{Croft_PRB_2014,Thampy_PRB_2014,Wen_NC_2019,Miao_nQM_2021,Arx_nQM_2023} which focus on the evolution of the charge correlations and show that they weaken with doping. Our approach contrasts with these studies in that we collect systematic wavevector-dependent RIXS data with high energy resolution and at both Cu-$L_3$ and O-$K$ edges and fit the data to two (Cu-$L_3$) or four (O-$K$) phonon modes and a (quasi-)elastic peak.  We interpret changes of the phonon intensity and dispersion anomalies as signaling the presence of CDFs.  Using this method we distinguish three contributions to the charge response which change differently with doping. 

The article is organized as follows. The experimental methods including sample growth, characterization, the RIXS method and fitting are detailed in Sec.~\ref{Sec:experimental_method}. The results including RIXS data and the extracted parameters are described in \ref{Sec:results}. In Sec.~\ref{Sec:evolution} we described how the charge order and charge density fluctuations evolve with doping. In Sec.~\ref{Sec:other_experiments} we discuss how our results relate to other experiments.  In Sec.~\ref{Sec:discussion} we discuss the implication of our observations for models of the cuprate phase diagram and the transport properties. Finally, in Sec.~\ref{Sec:sum} we summarize the main conclusion of the work.

\section{Experimental Method}
\label{Sec:experimental_method}
\subsection{Sample growth and characterization}\label{sec:sample}

We studied single crystals of La$_{2-x}$Sr$_x$CuO$_4$ with Sr doping values ranging from $x=0.11$ to $x=0.22$, grown by the traveling solvent floating zone method, followed by annealing in flowing oxygen at 800$^\circ$C for 2 weeks. The Sr content $x$, which we take as to be equivalent to the hole doping $p$, was measured by electron probe micro-analysis (EPMA).  Pieces used in the current study were cut from high-quality single crystal growths that have been used in previous inelastic neutron scattering (INS), X-ray and transport studies~\cite{Vignolle_NP_2007,Croft_PRB_2014,Cooper2009_CWV}. $T_c$ was determined by the midpoint of the magnetic susceptibility transition. The measured Sr doping and $T_c$ values of the samples used in this study are listed in Table.~\ref{tab:samples}. Samples were post cleaved in the load lock in a vacuum of $1\times10^{-8}$\;mbar or better. Samples were oriented \textit{in-situ} on the (002) reflection, combined with alignment on the CDW peak, which orders along the primary crystallographic directions or using information from \textit{ex-situ} Laue measurements.

\begin{table}[!htb]
\begin{tabular}{|c|c|}
\hline
Sr concentration, $x$ & Midpoint, $T_c$ (K) \\ \hline
0.110(2)           & 24.2(5)  \\
0.125(5)              & 31.0(5)   \\
0.160(2)               & 38.4(5)   \\
0.19(1)               & 33.5(5)   \\
0.215(5)               & 26.0(5)  \\ \hline

\end{tabular}
\caption{Sample $T_c$ values measured as the midpoint of the susceptibility transition, and Sr doping concentration values, determined by EPMA.}
\label{tab:samples}
\end{table}

\begin{figure*}[htb!]
\includegraphics[]{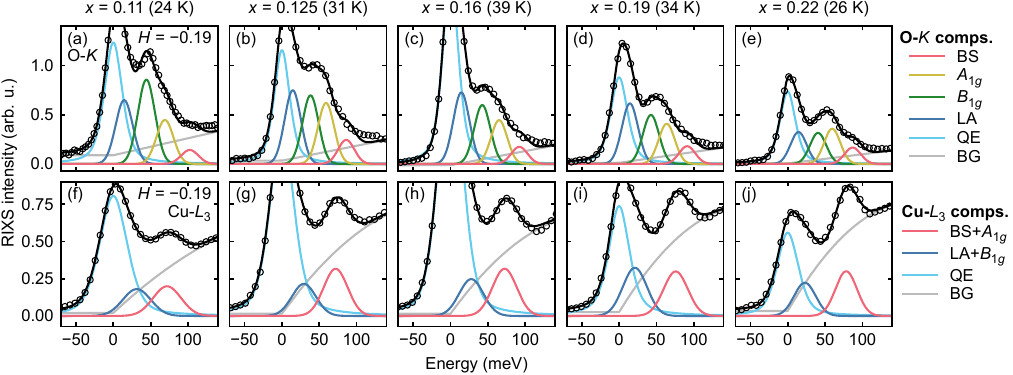}
\caption{\textbf{RIXS fitting scheme.} (a)-(e) Five-peak fitting scheme for O-$K$ RIXS fits for the five doping values. The overall fit is plotted in black, and the quasi-elastic (QE), longitudinal acoustic (LA) mode, $A_{1g}$ and $B_{1g}$ bond-buckling modes, bond-stretching (BS) phonon and polynomial background (BG) components are plotted. (f)-(j) Three-peak fitting scheme for Cu-$L_3$ RIXS fits. The overall fit is plotted in black, and the quasi-elastic, $p_1$, a combination of the longitudinal acoustic + $B_{1g}$ mode (LA + $B_{1g}$), and $p_2$, a combination of bond-stretching + $A_{1g}$ mode (BS + $A_{1g}$), and polynomial background (BG) components are plotted. All spectra are taken at $H=-0.19$.}
\label{fig:fits} 

\end{figure*}
\subsection{Spectrometer details}\label{sec:spectrometer}
RIXS measurements were taken at beamline I21, Diamond Light Source~\cite{Zhou_JoSR_2022}. In this article we use both Cu-$L_3$ and O-$K$ edge RIXS to measure the CDW and associated charge fluctuations in LSCO. Momentum transfer is labeled in reciprocal lattice units (r.l.u.) of the tetragonal unit cell, where $\mathbf{Q} = H \mathbf{a}^{\star} + K \mathbf{b}^{\star} + L \mathbf{c}^{\star}$, with $a=b=3.86$\;\AA\ and $c=13.1$\;\AA. The photon energy loss is $E=E_i-E_f=\hbar\omega$, where $E_i$ and $E_f$ are the initial and final photon energies.  Scans were taken with grazing incidence (i.e. $\theta < \Omega$), where $\theta$ is the incidence angle and $\Omega$ is the scattering angle. Grazing incidence is denoted with negative $H$. We used linear vertical polarized X-rays for all measurements. For Cu-$L_3$ ($E_i\approx 928\;$eV) the instrumental resolution (full-width-at-half-maximum) $\Delta E_{\mathrm{FWHM}} \approx 37\;$meV and for the O-$K$ ($E_i\approx527\;$eV) edge with $\Delta E_{\mathrm{FWHM}} \approx 27\;$meV. The CDW in LSCO is weakly correlated in $L$, but is peaked at integer + $1/2$ positions~\cite{Croft_PRB_2014}. 

The RIXS $H$ scans were taken at fixed scattering angle values of $\Omega=117^\circ$ and $\Omega=154^\circ$, to give $L=1.5$ and $L=0.7$ (r.l.u.) at the expected in-plane CDW wavevector $H=0.23$ for the Cu-$L_3$ and O-$K$ edge measurements respectively. With these constraints on the scattering angle and the low incident energy of soft X-rays, the highest $H$ value that can be reached at Cu-$L_3$ is $\left| H \right |=0.38$, and at O-$K$ is $\left| H \right |=0.3$. Measurements were normalized to the orbital excitations in range $[1,3]\;$eV, for both O-$K$ and Cu-$L_3$ data, to allow for comparisons between different doping values. Measurements were performed at the superconducting transition temperature $T_c$ of the respective samples, which are listed in Table.~\ref{tab:samples}.

\subsection{RIXS Measurements} \label{sec:rixs}

Undoped cuprates are classified as charge-transfer Mott insulators~\cite{Zaanen_PRL_1985}, since the Mott gap is determined by the anion charge transfer $\Delta$, rather than the onsite Coulomb repulsion $U$. Due to strong hybridization between Cu $3d$ and O $2p$, the valence band has oxygen character and can be understood as a single band, called the Zhang-Rice singlet (ZRS)~\cite{Zhang_PRB_1988} [see Fig.~\ref{fig:overview}(a)]. The Cu-$L$ RIXS involves excitations between the O $2p$ level and the Cu $d^{10}$ upper Hubbard band (UHB).   While the O-$K$ RIXS involves excitations from the O $1s$ state to the ZRS.  The ZRS appears as a pre-peak  in O-$K$ XAS [see Fig.~\ref{fig:overview}(b)], at lower energy than the UHB, and is where doped holes are seen to reside~\cite{Chen_PRL_1991}.

For the O-$K$ RIXS measurements, we tuned the incident X-ray energy to the ZRS pre-peak in O-$K$ XAS, plotted in Fig.~\ref{fig:overview}b, at around 527.3\;eV. An O-$K$ RIXS spectrum taken at $H=-0.29$ of $x=0.19$ LSCO is seen in Fig.~\ref{fig:overview}c. From right to left, we see a broad orbital excitation centered at 2500\;meV. Next, we see a very broad, non-dispersing bimagnon centered at around 600\;meV, and finally several low-energy features below 100\;meV.

For the Cu-$L_3$ RIXS measurements, the incident X-ray energy was tuned to the maximum of the Cu-$L_3$ $d^9$ absorption peak, at about $E_i=928$\;eV, plotted in Fig.~\ref{fig:overview}(d). A Cu-$L_3$ RIXS spectrum taken at $H=-0.29$ of $x=0.19$ LSCO is seen in Fig.~\ref{fig:overview}(e). From right to left, we see $dd$ crystal field excitations centered at around 1600\;meV energy loss, a dispersing paramagnon centered at around 300\;meV, and again several low-energy excitations below 100\;meV. 

\begin{figure*}[htb!]
\includegraphics[]{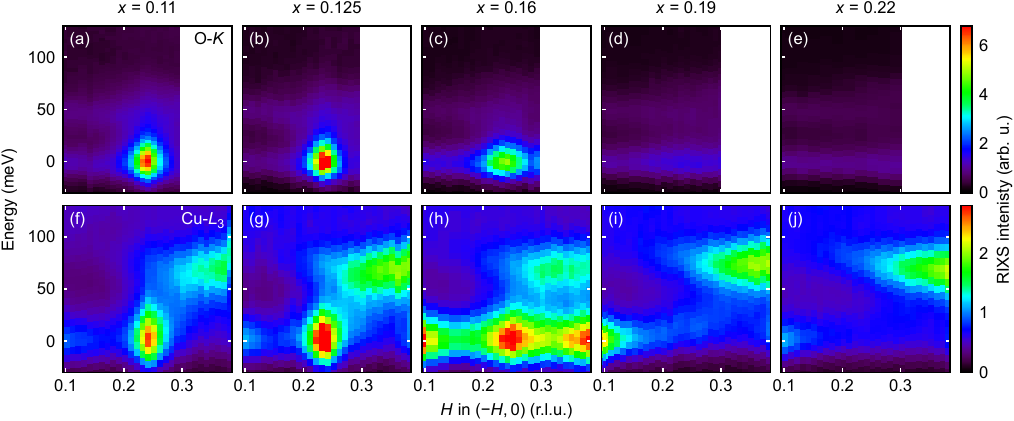}
\caption{\textbf{$H$-dependent RIXS color maps.} (a)-(e) Low-energy O-$K$ RIXS color maps for samples with different Sr doping levels. (f)-(j) Low-energy Cu-$L_3$ RIXS color maps. The $x=0.16$ sample used to collect data in this paper showed a larger elastic background [panels (c) and (h)] than the other samples. The background is most likely from poor surface quality resulting from a poor cleave.}
\label{fig:colour} 
\end{figure*}

Fig.~\ref{fig:fits} shows RIXS spectra collected at the O-$K$ and Cu-$L_3$ edges for each of the five compositions investigated and $H=-0.29$. Fig.~\ref{fig:colour} shows collections of spectra such as those in Fig.~\ref{fig:fits} plotted as $H$-$E$ color intensity maps where $E=\hbar\omega$ is the energy loss.

\subsection{Interpretation and Fitting}  
\label{sec:fitting_models}

Measurements of charge excitations using RIXS, such as those presented here are interpreted in terms of the dynamical structure factor $S(\mathbf{Q},\omega)$ defined as the Fourier transform of the charge density-density correlation function $\left<n(\mathbf{r},t)n(0,0)\right>$ in time and space \cite{Ament_RoMP_2011}. Further, $S(\mathbf{Q},\omega)$ can be related to a generalized charge susceptibility $\chi(\mathbf{Q},\omega)$ through the fluctuation-dissipation theorem $S(\mathbf{Q},\omega)=2 \hbar (1-e^{-\beta\hbar\omega})^{-1}\chi^{\prime\prime}(\mathbf{Q},\omega)$. The actual RIXS cross section will also contain other factors and matrix elements determined by the polarization of the X-rays, orbital character and geometry\cite{Ament_RoMP_2011}.  A diagrammatic approach \cite{Devereaux_PRX_2016, Huang_PRX_2021} shows that the coupling between CDFs and phonons will lead to peaks in the RIXS intensity, $I_{\text{RIXS}}$, where phonons cross CDFs (see schematic in Fig.~\ref{fig:schematic}). This picture is supported by RIXS data in the cuprates where a number of oxygen phonons with strongly modulated intensities \cite{Chaix_NP_2017,Huang_PRX_2021} are observed at the Cu-$L$ and O-$K$ edges.    
Certain phonons involving motion of the in-plane oxygen atoms are visible in RIXS because of their strong electron-phonon coupling\cite{Devereaux_PRX_2016}.  These are the longitudinal acoustic (LA), the $A_{1g}$ and $B_{1g}$ $c$-axis modes, and bond-stretching (BS) mode. This picture is supported by very-high resolution ($\Delta E_{\mathrm{FWHM}}=16$~meV) O-$K$ RIXS measurements~\cite{Huang_PRX_2021} on LSCO ($x=0.15$) which are fitted to four inelastic components.  The analysis in this paper follows this work.  With the energy resolution of the data currently presented, we are able to understand the RIXS spectra in terms of these four phonons at O-$K$, or two phonons at Cu-$L_3$.  

\subsubsection{Five-Peak Model (O-$K$ edge)}
\label{Sec:5_peak_model}
We fit the low-energy part of the O-$K$ RIXS spectra in the range $[-100, 200]\;$meV, to a five-peak model. This consists of a pseudovoigt (pV) function (a linear combination of Gaussian and Lorentzian functions with $\mathrm{pV}(0)=1$) to fit the quasi-elastic (QE, $\omega=0$) scattering, Gaussians with the experimental resolution width to fit four phonon modes, and a polynomial background. From low to high energy [see Fig.~\ref{fig:phonons}(p)], the phonon modes are a $\sim 14.5\;$meV longitudinal acoustic (LA) mode, a $\sim 39\;$meV $A_{1g}$ buckling mode, a $\sim 56\;$meV $B_{1g}$ buckling mode, and a $\sim$80\;meV bond stretching (BS) phonon~\cite{Huang_PRX_2021}. For a cut along the trajectory $\mathbf{Q}=(H,0)$, the fitted intensity can be written as  

\begin{equation}
\begin{split}
    I_{\text{RIXS}}(H,\omega) = & I^{\mathrm{QE}}(H) \;\mathrm{pV}(\omega) + \\\sum_{i=\{\mathrm{LA}, B_{1g}, A_{1g}, \mathrm{BS}\}} &
     I^i(H) \; \exp\left[{-\frac{(\omega - \omega_i)^2}{2 \sigma^2_\mathrm{res}} } \right]  + \mathrm{B.G.}, \label{eqn:5-peak}
\end{split}
\end{equation}
where $I_{\text{RIXS}}(H,\omega)$ is the RIXS intensity, $i$ indexes the four phonon peaks, $I^i$ is the intensity of the $i$th peak and, $\omega_i$ is the peak position, and the instrumental resolution $\Delta E_{\textrm{FWHM}}$ $\approx$ 27\;meV $\approx 2.335\sigma_\mathrm{res}$.

The current experimental resolution is not sufficient to allow the LA phonon frequency to vary in the fitting procedure, so it is fixed to $14.5\;$meV, which is the frequency measured in LSCO at $(0.25, 0, 0)$~\cite{Pintschovius_PCS_1991} by INS and is consistent with a previous RIXS study~\cite{Huang_PRX_2021}. When the frequencies were allowed to vary, the frequencies of the $A_{1g}$ and $B_{1g}$ buckling modes are consistent with Ref.~\onlinecite{Huang_PRX_2021}. However, when we examine the $H$ dependence of the intensity, the frequencies are fixed to their average values of 39\;meV and 56\;meV respectively. Example fits for $H = -0.19$ are plotted in Fig.~\ref{fig:fits}(a-e) and the result of fitting the data displayed in Fig.~\ref{fig:colour} is shown in Fig.~\ref{fig:phonons}.

\subsubsection{Three-Peak Model (Cu-$L$)}
\label{Sec:3_peak_model}
To fit the low-energy $[-100, 200]\;$meV Cu-$L_3$ spectra, a three-peak model was used. It consisted of a pseudovoigt peak to fit quasi-elastic scattering, and two resolution limited Gaussians to fit $\sim 30$ and $\sim 80\;$meV modes. This can be written as 

\begin{equation}
\begin{split}
    I_{\text{RIXS}}(H,\omega) = & I^{\mathrm{QE}}(H) \;\mathrm{pV}(\omega) + \\ \sum_{i=\{p_1, p_2\}} 
    & I^i(H) \; \exp\left[{-\frac{(\omega - \omega_i)^2}{2 \sigma^2_\mathrm{res}} } \right] + \mathrm{B.G.},
\end{split}
\end{equation}
where $i$ indexes the two identifiable (combined) phonon modes $p_1$ and $p_2$ and $\Delta_{\mathrm{FWHM}} \approx2.335 \sigma_\mathrm{res} \approx 37\;$meV.

\begin{figure*}[htb!]
\includegraphics[]{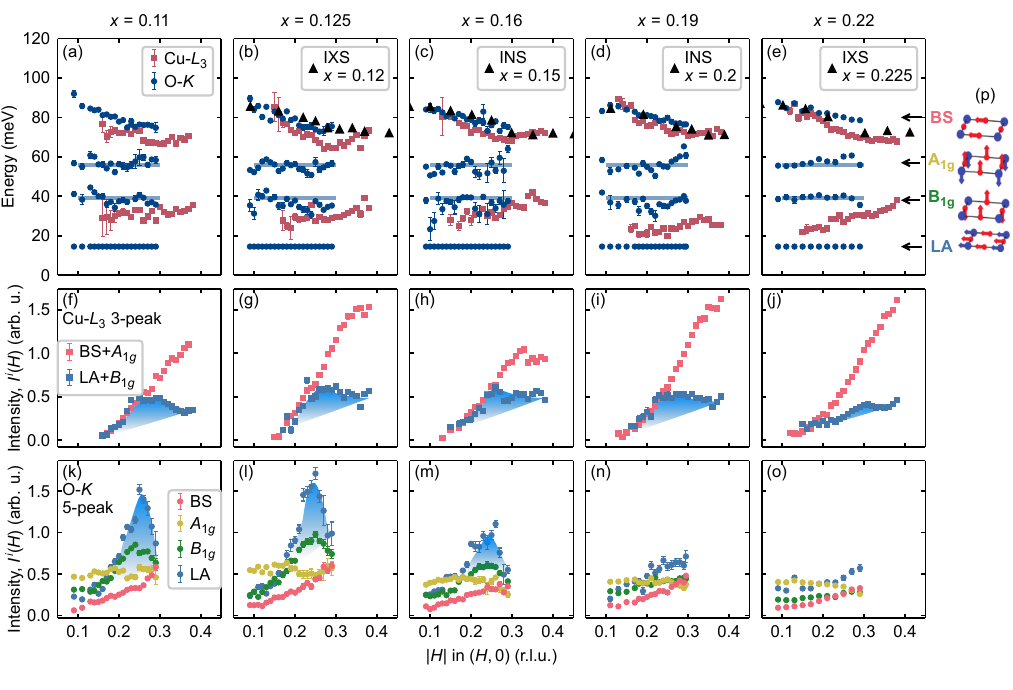}
\caption{\textbf{Phonons measured with RIXS in LSCO.} (a)-(e) Fitted phonon dispersion, measured with both O-$K$ (blue markers) and Cu-$L_3$ (red markers) RIXS. The black points are the dispersion of the BS phonon from samples with similar doping levels, measured with IXS/INS, with data taken from Refs.~\onlinecite{Fukuda_PRB_2005,Park_PRB_2014}. The pale blue lines mark the average value of the dispersion. (f)-(g) Fitted Cu-$L_3$ phonon intensities (Gaussian height), using the three-peak model described in the text. (k)-(o) Fitted O-$K$ phonon intensities, using the five-peak model described in the text. Blue shaded areas are the approximate signal due to CDFs in the LA data. (p) Illustration of the BS, $A_{1g}$, $B_{1g}$ and LA phonons modes in the CuO$_2$ planes.} 
\label{fig:phonons} 
\end{figure*}

Given the phonon energies discussed in Sec.~\ref{Sec:5_peak_model} for the O-$K$ data, $p_1$, the $\sim 30\;$meV mode is likely a mixture of the LA and $B_{1g}$ buckling phonons, while $p_2$, the 80\;meV mode, is likely a mixture of the $A_{1g}$ buckling and the bond-stretching (BS) phonon present in cuprates~\cite{Chaix_NP_2017,Lin_PRL_2020}. Example fits for $H = -0.29$ are plotted in Fig.~\ref{fig:fits}(f-j) and the $H$-dependent results (phonon energies and peak intensities) are displayed in Fig.~\ref{fig:phonons}. While a two-peak fit has been commonly used in the literature (e.g. Refs.~\onlinecite{Lin_PRL_2020,Lee_NP_2020}), adding a third significantly increases the fit quality, especially near $Q_\mathrm{CDW}$. Three peak fits are used in more recent work, e.g. Refs.~\onlinecite{Arpaia_NC_2023,Zou_NC_2024}. Additionally, with the three-peak model, the energy of the BS phonon [Fig.~\ref{fig:phonons}(a-e)] more closely resembles that measured with INS and inelastic X-ray scattering (IXS) (see e.g. Refs.~\onlinecite{Fukuda_PRB_2005,Park_PRB_2014}), while with a two-peak fit, it is around 10\;meV lower than IXS and INS~\cite{Lin_PRL_2020}.  

\begin{figure*}[t]
\includegraphics[]{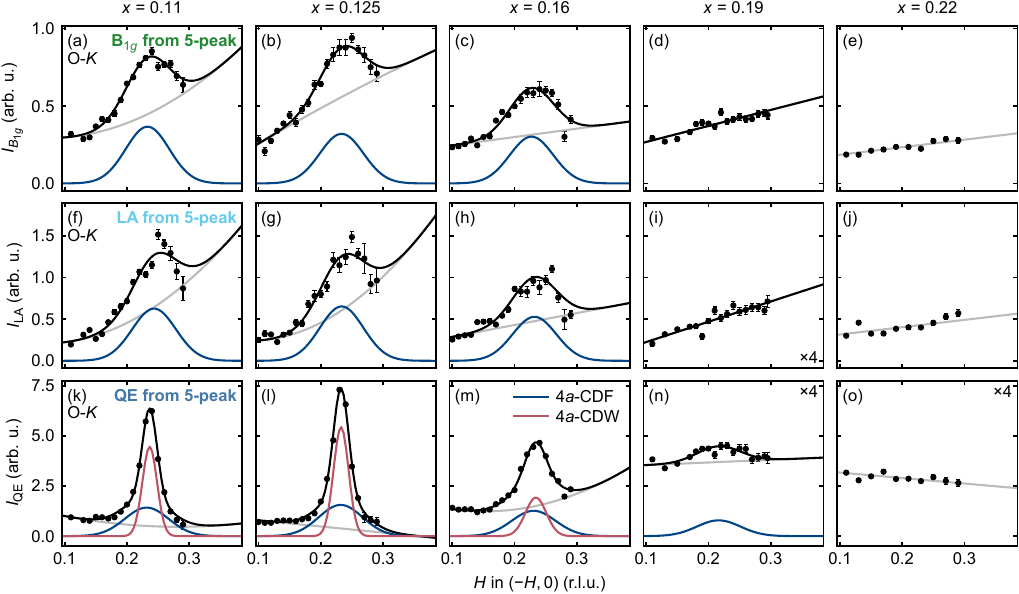}
\caption{\textbf{Charge density fluctuations (CDFs) obtained from five-peak fit to phonons measured by O-$K$ RIXS.} $H$ dependence of fitted phonon intensities, obtained from fitting the O-$K$ RIXS spectra using the five-peak model described in the text. (a)-(e) $B_{1g}$ phonons, (f)-(j) LA phonons. The $H$ dependence of the quasi-elastic peak is plotted in (k)-(o). The total fit is plotted in black, with Gaussian components in blue or red, and the background is gray. Intensities in panels (i), (n) and (o) have been quadrupled for visibility}
\label{fig:OK_2D_fits} 
\end{figure*}

\section{Results}
\label{Sec:results}
\label{sec:colourmap}
Color intensity $I_{\text{RIXS}}(H,\omega)$ maps of O-$K$ and Cu-$L_3$ RIXS scattering are shown in Fig.~\ref{fig:colour}. The most notable feature, for $x \le 0.16$, is the strong scattering near $|H|=\delta$ and $E \approx 0$. This is due to the CDW and CDF scattering. A second strong feature in the Cu-$L_3$ RIXS is observed near $E \approx 70$~meV and $H \approx 0.4$ and is due to the oxygen bond-stretching phonon, while the O-$K$ RIXS couples most strongly to a $\sim 50\;$meV oxygen bond-buckling mode.

\subsection{$E$-dependent fits of phonons}
\label{Sec:phonon_fits}

The models described in Secs.~\ref{Sec:5_peak_model} and \ref{Sec:3_peak_model} were used to fit the phonon dispersions and intensities of the phonons. Example fits are shown in Fig.~\ref{fig:fits}.
In Figs.~\ref{fig:phonons}(a-e) we plot the fitted phonon dispersions at both O-$K$ and Cu-$L_3$ resonances
for all measured samples. Here we will describe the dispersion and the intensity variation of the modes phonon by phonon, starting from the BS phonon and going down in energy.

\subsubsection{BS phonon}

The BS mode is detected with both O-$K$ and Cu-$L_3$ RIXS. Starting from the top of Figs.~\ref{fig:phonons}(a-e), we see generally good agreement between O-$K$ and Cu-$L_3$ with $\sim 80\;$meV BS phonon. We have also plotted IXS and INS measurements of the BS phonon on similar sample compositions from Refs.~\onlinecite{Fukuda_PRB_2005,Park_PRB_2014}. We see there is good agreement for the overdoped samples. For the RIXS data there appears to be a softening near $|H| \approx 0.3$ for $x=0.11$, 0.125 and 0.16 [Fig.~\ref{fig:phonons}(a-c)] suggesting the presence of charge excitations. 

The fitted $p_2$ phonon intensity (that is, the height of the fitted resolution-limited Gaussian) of the BS + $A_{1g}$ mode measured at Cu-$L_3$ are as red markers in Figs.~\ref{fig:phonons}(f-j), while the O-$K$ BS phonon intensities are plotted as red markers in Figs.~\ref{fig:phonons}(k-o). The mode intensity measured at both edges shows an increasing intensity as $H$ increases. This qualitatively fits with the $\sim \sin^2(\pi H)$ behavior of momentum dependence of the electron-phonon coupling, which the BS phonon in cuprates has been shown to obey~\cite{Devereaux_PRX_2016}.

\subsubsection{$A_{1g}$ and $B_{1g}$}


The $A_{1g}$ and $B_{1g}$ buckling modes can be separated with the O-$K$ RIXS and the energy dispersions are plotted in Figs.~\ref{fig:phonons}(a-e). They show little dispersion with average energies of around 56\;meV and 39\;meV respectively. The dispersion shows similar behavior between samples.

The O-$K$ $A_{1g}$ and $B_{1g}$ phonon intensities are plotted as yellow and green markers respectively in Figs.~\ref{fig:phonons}(k-o). The $A_{1g}$ mode shows a roughly constant intensity and the $B_{1g}$ mode (green) shows a peak near $|H| \approx 0.24$ for $x=0.11$, 0.125 and 0.16 indicating the presence of CDFs with $|H| \approx 0.24$ for these compositions.   For the Cu-$L$ data, the poorer resolution means that we see a combination of the LA and $B_{1g}$ modes (which we have denoted $p_1$).  The (LA+$B_{1g}$) Cu-$L$ data in Figs.~\ref{fig:phonons}(f-j) shows a CDF contribution near $|H| \approx 0.24$, also seen at the O-$K$ edge,  $x=0.11$, 0.125 and 0.16. The data also suggest there are CDFs near $|H| \approx 0.3$ (see also Refs.~\onlinecite{Fukuda_PRB_2005,Park_PRB_2014,Arx_nQM_2023}) for all compositions studied. These two contributions are highlighted by the shading in Figs.~\ref{fig:phonons}(f-j).

\subsubsection{LA phonon}

The lowest energy mode in the five-peak model used to fit the O-$K$ data is the LA phonon, which we have fixed to 14.5\;meV. It is plotted as blue markers in Figs.~\ref{fig:phonons}(k-o). Within the measured range of $0.1 < H < 0.3$, it has a weak dispersion that is consistent with other INS~\cite{Pintschovius_PCS_1991} and RIXS~\cite{Pintschovius_PCS_1991,Huang_PRX_2021} measurements. 
At O-$K$, the LA phonon mode has a peak intensity due to CDFs at $H \approx -0.24$ for $0.11 \leq x \leq 0.19$.  The peak is largest at $x=0.125$.

In summary, our analysis demonstrates that that the spectra can be largely understood in terms of  coupling between CDFs and phonons  \cite{Devereaux_PRX_2016,Huang_PRX_2021}. The dynamic CDFs have two contributions: one peaked at $|H| \approx 0.24$ and a second peaked at $|H| \approx 0.3$. 

\begin{figure*}[htb!]
\includegraphics[]{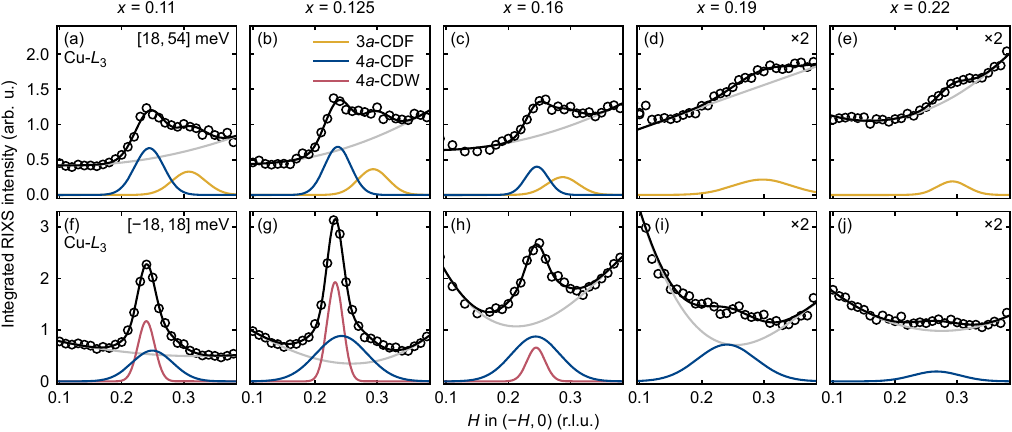}
\caption{\textbf{Charge density fluctuations measured with Cu-$L_3$ RIXS.} (a)-(e) $[18,54]\;$meV Cu-$L_3$ RIXS cut. (f)-(j) Cu-$L_3$ quasi-elastic RIXS cut, obtained by integrating RIXS in $[-18,18]\;$meV. Data have been fit to either a sum of two Gaussians and a polynomial background or one Gaussian and a polynomial background. The total fit is plotted in a black line, and fit components are plotted in colored lines. Intensities in (d), (e), (i), and (j) have been doubled for visibility.}
\label{fig:cu-intensity} 
\end{figure*}

\subsection{CDW and CDF at the O-$K$ resonance}
\label{sec:ok-cdw}
  
As discussed in Sec.~\ref{sec:intro}, previous studies \cite{Ghiringhelli_S_2012,Croft_PRB_2014,Arpaia_S_2019,Arx_nQM_2023} have highlighted the multi-component nature of the charge correlations in cuprates. Our O-$K$ data with their good energy resolution contain information about the energy scales of the fluctuations associated with the individual components.  Signatures of various components are contained in the the $H$-dependence of the phonon intensities $I^i(H)$, $i=\{\mathrm{LA}, B_{1g}, A_{1g}, \mathrm{BS}\}$ [see Eqn.~\ref{eqn:5-peak} and Figs.~\ref{fig:phonons}(k-o)].  

Our data suggests there are three components to the underlying charge correlations: (i) a \textit{quasielastic} CDW contribution ($4a$-CDW) peaked at $H=\delta_{4a}\approx 0.24$ and $\omega=0$; (ii) a broader CDF \textit{inelastic} component ($4a$-CDF) peaked at $H=\delta_{4a}\approx 0.24$; and (iii) a  broader CDF \textit{inelastic} component ($3a$-CDF) peaked at $H=\delta_{3a}\approx 0.3$ (only accessible in $H$ at the Cu-$L$ resonance). We have denoted these as $3a/4a$ simply for ease of naming, because the period of the components is roughly 3 or 4 lattice units. The inelastic $3a$-CDF and $4a$-CDF components are not seen in  $I^{A_{1g}}$ or $I^{\text{BS}}$ suggesting the energy scale of these components is less than 50 meV.     

Due to the finite instrumental resolution, $I^{\mathrm{QE}}(H)$ contains contributions from the CDW and CDF components. We model these as
\begin{align}
I^{\mathrm{QE}}(H) &= I_{4a\text{-CDW}} \; \exp\left[{-\frac{(H-\delta_{4a})^2}{2 \sigma^2_{4a\text{-CDW}}} } \right] \nonumber \\ + I_{4a\text{-CDF}}(\omega=0) & 
\; \exp \left[{-\frac{(H-\delta_{4a})^2}{2 \sigma^2_{4a\text{-CDF}}} } \right] +\text{B.G.} \label{eqn:phenom_1},
\end{align}
where $I_{4a\text{-CDW}}$ is the intensity of the $4a\text{-CDW}$ component and $I_{4a\text{-CDF}}(\omega=0)$ is the intensity of the CDF component.  The higher energy data are modeled as
\begin{align}
I^i(H) &= I_{4a\text{-CDF}}(\omega_i)  
\; \exp \left[{-\frac{(H-\delta_{4a})^2}{2 \sigma^2_{4a\text{-CDF}}} } \right] \nonumber \\ + I_{3a\text{-CDF}}(\omega_i) & \; \exp \left[{-\frac{(H-\delta_{3a})^2}{2 \sigma^2_{3a\text{-CDF}}} }\right] 
+ \text{B.G.},
\label{eqn:phenom_2}
\end{align}
where $I_{4a\text{-CDF}}(\omega_i)$ is measured at a phonon with frequency $\omega_i$ and the $I_{3a\text{-CDF}}(\omega_i)$ component can only be measured at the Cu-$L_3$ edge. 

If we consider the $x=0.125$ sample as an example of our data. Fig.~\ref{fig:OK_2D_fits}(l) shows a peak near $|H| \approx 0.24$ for $I^{\mathrm{QE}}(H)$. Peaks are also seen for the LA and $B_{1g}$ phonons in Fig.~\ref{fig:phonons}(l) and ~\ref{fig:OK_2D_fits}(a,f), but not obviously for the $A_{1g}$ and BS modes. The peak width of the LA and $B_{1g}$ phonons is larger than that of the quasi-elastic ($E=0$) scattering.
Thus, our O-$K$ edge data can be understood using the first two terms of Eqn.~\ref{eqn:phenom_1} representing a narrower QE CDW component with $\sigma_{4a\mathrm{-CDW}} \sim 0.01\;$r.l.u. ($\xi \sim 35\;\mathrm{\AA}$) and a broader inelastic CDF component seen mostly in the LA and $B_{1g}$ phonons, with $\sigma_{4a\mathrm{-CDF}} \sim 0.04\;$r.l.u. ($\xi \sim 15\;\mathrm{\AA}$). 

The result of fitting Eqn.~\ref{eqn:phenom_1}-\ref{eqn:phenom_2} to the data in Figs.~\ref{fig:phonons} and \ref{fig:OK_2D_fits} is shown in Fig.~\ref{fig:phase}(c-e). The narrow $4a$-CDW QE component is only observed from $x=0.11$ to $x=0.16$. The $x=0.16$ quasi-elastic is best fit to a combination of both components, Fig.~\ref{fig:OK_2D_fits}(m). This was done by fixing the width of the broad component at the average value of $\sigma = 0.04\;$r.l.u. and allowing the narrow component to vary. This is to reduce the number of fitting parameters when the CDW contribution is weak. The $4a$-CDF is no longer detected in the $x=0.22$ quasi-elastic channel. 

\subsection{CDW and CDF at Cu-$L_3$ resonance}
\label{sec:cu-cdw}

RIXS measurements at the Cu-$L_3$ resonance provide complementary information to those made at the O-$K$ resonance: they are sensitive to phonons in a different way, and the higher energy of the Cu-$L_3$ resonance means that measurements can be made to larger $|H| \approx 0.4$.   
However, the energy resolution $\Delta E \approx 37$\;meV is worse than at the O-$K$ resonance; therefore, the phonons cannot be isolated in the same way as for the O-$K$ data. For simplicity, we integrate our data over an energy range corresponding to the experimental resolution and fit to the resulting points.  Fig.~\ref{fig:cu-intensity} shows $H$-dependent cuts integrated over energy ranges $[-18, 18]$\;meV, and $[18,54]\;$meV. The cuts are consistent with those obtained by the 3-peak fits shown in Fig.~\ref{fig:phonons}(f-j). 

The Cu-$L_3$ quasi-elastic $E \in [-18,18]\;$meV cuts for the $x=0.11$, 0.125 and 0.16 can be fitted to the sum of Gaussian peaks and a 2nd order polynomial background [Figs.~\ref{fig:cu-intensity}(f,g)]. As with the 2D O-$K$ fits, one peak is broader than the other ($\sigma_{\text{CDF}} \sim 0.04$~r.l.u compared to $\sigma_{\text{CDW}} \sim 0.01$~r.l.u.). Again, we identify the broad and narrow components centered at $H\approx -0.24$, with the $4a$-CDF and $4a$-CDW contributions respectively. For $x=0.16$, we fix $\sigma_{\text{CDF}} = 0.04$~r.l.u, as with the O-$K$ data. For $x=0.19$ we fit one broad peak [Figs.~\ref{fig:cu-intensity}(h,i)]. A weak peak centered at $H\approx-0.27$ is seen for $x=0.22$ [Fig.~\ref{fig:cu-intensity}(j)]. A summary of the fitted values for the quasi-elastic cut are plotted in Fig.~\ref{fig:phase}(f)-(h).

The finite energy $E \in [18,54]\;$meV cut at Cu-$L_3$ which probes the CDF component is plotted in Figs.~\ref{fig:cu-intensity}(a-e). It is possible to access a wider range in $H$ at the Cu-$L_3$ resonance compared to O-$K$, we observe scattering centered at $H\approx -0.29$ in the $x=0.22$ sample. For $x\leq 0.16$, there is a shoulder of scattering in addition to the $H\approx-0.24$ scattering from the $4a$ CDW/CDF. Here we choose to interpret the scattering as the sum of two peaks with at two distinct $H$ values.

We apply a two-Gaussian fit to the $0.11 \leq x \leq 0.19$ $[18,54]\;$meV cut. Using what we know from the O-$K$ data, that the center of the $4a$ CDF is similar to the $4a$-CDW, we fix one of the CDF peaks to the center of the $H\approx -0.24$ CDW measured in the quasi-elastic cut. The other peak was allowed to float. The result is a second peak centered at $H\approx-0.29$ in all the measured samples -- we call this ``$3a$-CDF'' since $\lambda \approx a/0.29 \approx 3.4a$. The $3a$ CDF peak has a width $\sigma_{3a\text{-CDF}} \approx 0.04\;$r.l.u. which is comparable with $\sigma_{4a\text{-CDF}}$ and approximately constant with doping $x$. A summary of the fitted values for the inelastic cuts is plotted in Fig.~\ref{fig:phase}(i-k). Note that the $3a$-CDF can only be clearly observed at the Cu-$L$ resonance because the large energy allows $H>0.3$ to be easily reached. 

\section{Evolution of charge order and charge fluctuations with doping}\label{Sec:evolution}

\subsection{Three Components}
\label{sec:three_components}
\begin{figure*}[htb!]
\includegraphics[]{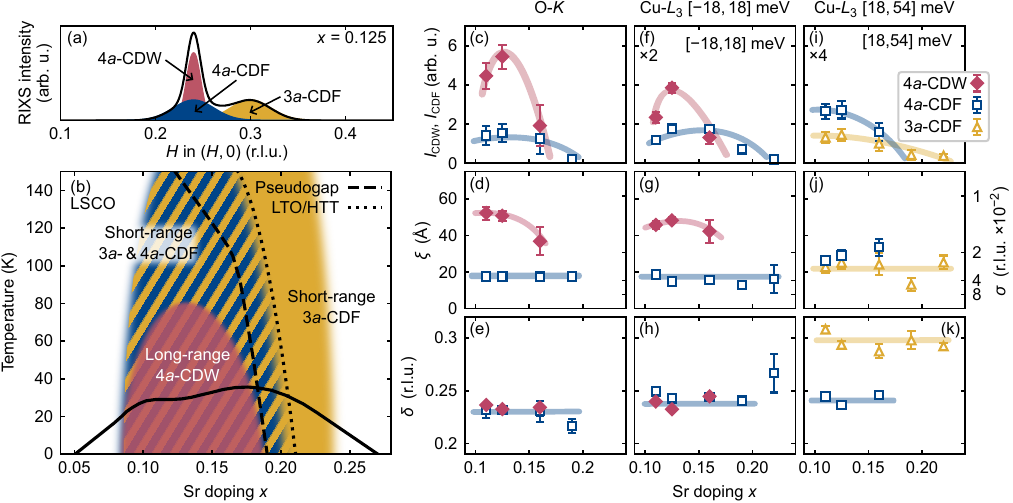}
\caption{\textbf{CDW phase diagram in LSCO.} (a) Schematic, in RIXS scattering intensity and $H$, of the three CDW components. (b) Schematic phase diagram, with regions where $4a$-CDW order (red), $4a$-CDF (blue) $3a$-CDF (yellow) are present. The superconducting dome is plotted in a black solid line. The pseudogap boundary is marked by the black dashed line, which is based on Nernst effect measurements~\cite{CyrChoiniere_PRB_2018}. The boundary between the low-temperature orthogonal (LTO) and high-temperature tetragonal phase (HTT), is marked with a black dotted line, based on data from Ref.~\onlinecite{Yamada_PRB_1998}. (c), (f), (i) Fitted peak intensities (Gaussian height); (d), (g), (j) correlation length $\xi$; and (e), (h), (k) CDW incommensurability $\delta$ for the 1D O-$K$ phonon fit, the 1D $[-18,18]\;$meV Cu-$L_3$ cut, and the the 1D $[18,54]\;$meV Cu-$L_3$ cut respectively. The secondary y axis in (d), (g), (j) is the fitted Gaussian width parameter $\sigma$ and is related to the correlation length by $\xi = 1/\sigma$. There are two $4a$-CDF markers at $x=0.16$ in (c)-(e) because there are both a broad QE component and an a broad LA fit present at this doping.}
\label{fig:phase} 
\end{figure*}

In this work, we have identified three components contributing to the charge correlations (see schematic in Fig.~\ref{fig:phase}(a) for how these appear in an $H$-scan). The first is longer-range $4a$-CDW order with $\mathbf{Q}=(\delta,0)$ and $\delta\approx0.24$.  This is elastic ($E=0$) within the resolution of the present experiment, however, this component may fluctuate on an energy scale $\Delta E \lesssim 1$~meV, as discussed in Refs.~\onlinecite{Blackburn_PRB_2013,Tacon_NP_2013}. Our measurements at both the O-$K$ and Cu-$L_3$ resonances suggest that the $4a$-CDW disappears at about $x=0.16$ and that for dopings $x=1/8$ (where it is strongest) it has a correlation length $\xi \approx 50$~\AA.  Our results are broadly in agreement with previous studies which report an increase below $\xi$ below $T_{\text{CDW}} \approx 80$~K signaling the onset of the CDW \cite{Croft_PRB_2014,Thampy_PRB_2014,Wen_NC_2019}. 
The second component ($4a$-CDF) corresponds to shorter-range correlations that also with  $\delta \approx 0.24$. In Sec.~\ref{sec:ok-cdw} we show that the $4a$-CDF component is dynamic and has the highest intensity for $\hbar\omega_{\text{LA}} \approx 14.5\;$meV. The third component ($3a$-CDF) is seen in the Cu-$L_3$ RIXS shown in Figs.~\ref{fig:cu-intensity}(a-e) is CDFs with $\delta\approx0.29$, that is seen in the energy range $[18, 54]\;$meV.

\subsection{``Phase diagram''}
Fig.~\ref{fig:phase}(b) shows a sketch of the regions of $x$-$T$ space where the three contributions to the charge correlations are strong.  This is based on  present and literature data. The $4a$-CDW component exists in the range $0.1 < x < 0.16$, is strongest at $x=0.125$, and shows a strong temperature dependence. The temperature where the CDW emerges is approximately 80\;K at $x=0.125$~\cite{Croft_PRB_2014,Thampy_PRB_2014,Wen_NC_2019,Miao_nQM_2021}. $4a$-CDFs in LSCO exists to at least 150\;K when measured with XRD~\cite{Croft_PRB_2014}, Cu-$L_3$ RIXS~\cite{Arx_nQM_2023} and 210\;K in O-$K$ RIXS~\cite{Huang_PRX_2021}. For $0.19< x < 0.22$, the $4a$-CDF is no longer detected in our data, while the $3a$-CDF, which may have a higher energy scale, is observed over the range $x \in [0.11,0.22]$. 

\subsection{Intensities, correlation lengths, and characteristic wavevectors}
Fig.~\ref{fig:phase}(c-k) shows the peak intensity (i.e. Gaussian height), correlation length $\xi$ and incommensurability $\delta$ describing the three components of the charge correlations described above.  We see [Fig.~\ref{fig:phase}(c-f)] that both the $4a$-CDW and $4a$-CDF, represented by red diamonds and blue squares, are strong at $x=0.125$. The $4a$-CDW component disappears between $x=0.16$ and 0.19, while the $4a$-CDF disappears between $x=0.19$ and 0.22. The $3a$-CDF, represented by yellow triangles, exists over all measured doping values, and extrapolation suggests that it would disappear at $x \approx 0.25$. $x=0.25$ is also the doping value where an anomaly in the BS phonon disappears. 

The fitted correlation lengths are plotted in Figs.~\ref{fig:phase}(d,g,j). The correlation lengths of the $3a$- and $4a$-CDFs are relatively doping independent, with similar values of around $\xi \approx 20$~\AA\ obtained for the two components. The correlation length of the $4a$-CDW, is maximized around $x=0.125$ at $\xi \approx 50$~\AA\, and it decreases slightly as doping is increased towards $x=0.16$. We find that our $\xi_{4a-\text{CDW}}$ is larger than that found \cite{Croft_PRB_2014} using hard X-rays ($\xi_{\text{X-ray}}\approx 30$~\AA), this is most likely due to the multi-component model used here which separates the CDW contribution of the CDFs. 

The CDW incommensurability $\delta$ of the three components is plotted in Figs.~\ref{fig:phase}(e,h,k). We find the $\delta$ for the $4a$ CDF and CDW (blue and red points) are constant with doping with $\delta=0.24$. The $3a$ CDF (yellow points) are centered consistently on $\delta=0.29$. 

\begin{figure*}[htb!]
\includegraphics[]{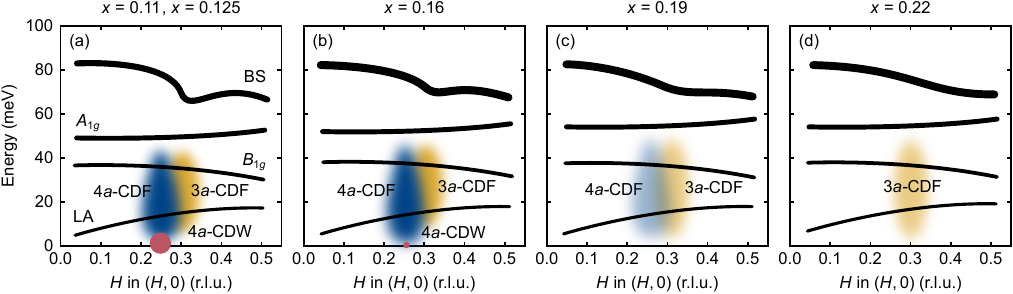}
\caption{\textbf{Schematic doping dependent charge fluctuations in LSCO.}  Doping dependent $(H,E)$ schematic of charge fluctuations. (a) Underdoped compositions, where $4a$-CDW is maximized (red circle), and the $3a$- and $4a$-CDF (yellow, blue regions) are detected. (b) In optimally doped compositions, the $4a$-CDW is reduced in intensity, and the $3a$- and $4a$-CDF are still present. (c) In overdoped compositions, $4a$-CDW is no longer present. (d) For the most overdoped composition only a weak $3a$-CDF remains.  }
\label{fig:schematic} 
\end{figure*}   

\subsection{Energy scales}
\label{sec:energy_scales}
A summary of the doping evolution of three components in $(H,E)$ of our observations is shown in  Fig.~\ref{fig:schematic}.  The $4a$-CDF component gives raise to intensity peaks in the 14.5\;meV LA and the 39\;meV $B_{1g}$ phonon modes, but is not seen significantly through the 56\;meV $A_{1g}$ mode. The $4a$-CDF gives intensity peaks in phonons of similar energies in the Cu-$L$ data.  However, the poorer resolution means that the LA and B$_{1g}$ phonons cannot be resolved.  

\begin{figure*}[htb!]
\includegraphics[]{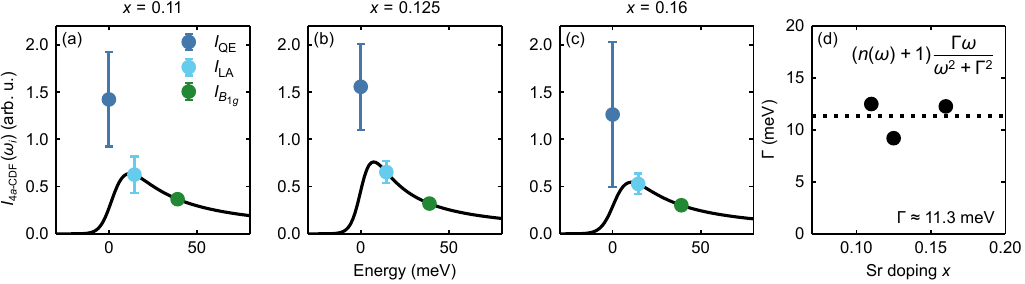}
\caption{\textbf{Energy scales of $4a$ CDFs.} (a)-(c) Fitted peak intensities from the 5-peak fitting procedure for the $4a$ CDFs. The non-zero $4a$ CDF intensities, i.e. the quasielastic (QE) and LA and $B_{1g}$ phonons, have been fitted to Lorentzian relaxation form given in Sec.~\ref{sec:energy_scales}. (d) The fitted relaxation rate $\Gamma$ for the three doping values, which has an average value of 11.3\;meV.}
\label{fig:energies} 
\end{figure*}

Fig.~\ref{fig:energies}(a-c) we plot the intensities $4a$-CDFs seen through the phonons together with the QE scattering measured at O-$K$ resonance.  If we assume that these intensities are proportional to the dynamic structure factor of the charge fluctuations $S(\mathbf{Q},\omega)$ then we can estimate the energy scale of the CDFs. In Fig.~\ref{fig:energies}(a-c) we fit a very overdamped harmonic oscillator response $\chi''(\mathbf{Q},\omega) \propto \omega\Gamma/(\omega^2+\Gamma^2)$ times a Bose factor (see Sec.~\ref{sec:fitting_models}) convolved with the instrumental resolution to the $I_{4a-\text{CDF}}(\omega_i)$ values.  The resulting fit yield a doping-independent relaxation rate $\hbar\Gamma \approx 11$~meV for $x \in [0.11,0.16]$, where there is a significant $4a$-CDF signal. 

\section{Relationship to other experiments}\label{Sec:other_experiments}

\subsection{RIXS on Cuprate Superconductors}

Our results agree broadly with other X-ray studies on LSCO\cite{Croft_PRB_2014, Wen_NC_2019,Lin_PRL_2020,Miao_nQM_2021,Arx_nQM_2023,Hong2025} which also reveal the multi-component nature of the charge correlations.   For example, a recent RIXS study of LSCO at the Cu-$L$ resonance \cite{Arx_nQM_2023} has characterized disappearance of the charge correlations with doping in terms of decreasing correlation length.  We conclude that the $4a$-CDW order disappears near $x \approx 0.16$, which is in agreement with this study \cite{Arx_nQM_2023} and in disagreement with Ref.~\onlinecite{Miao_nQM_2021} which suggests that the $4a$-CDW order persists to $x=0.21$.  Our work at the O-$K$ resonance contrasts with the Cu-$L$ RIXS\cite{Lin_PRL_2020,Arx_nQM_2023} and diffraction \cite{Croft_PRB_2014, Wen_NC_2019,Miao_nQM_2021} in that we are able to separate three components using good energy resolution showing that the broader component is inelastic.  Our work is agreement with very high-resolution O-$K$ RIXS experiments on LSCO ($x=0.15$) \cite{Huang_PRX_2021} which find evidence for RIXS excitations from due to coupling of CDW fluctuations to acoustic phonons peaked near 14~meV for $H=\delta$.

Signals of an inelastic CDFs are also seen with Cu-$L_3$ RIXS in Hg1210\cite{Yu_PRX_2020} and YBCO\cite{Arpaia_S_2019,Arpaia_NC_2023}. In YBCO\cite{Arpaia_NC_2023}, the wavevectors are different, however, at low temperatures CDFs are seen with $\hbar\Gamma = 10-15$~meV for $p=0.17-0.19$ i.e. comparable the values $\hbar\Gamma\approx11$~meV reported here for LSCO.

\subsection{Phonon anomalies in cuprates}
In LSCO, there is a softening and broadening of the 80\;meV BS phonon at $H\approx 0.3$~\cite{Fukuda_PRB_2005,Park_PRB_2014}. The phonon anomaly exists from around $0.05 \lesssim x \lesssim 0.25$. The anomaly is temperature independent up to 100\;K~\cite{Lin_PRL_2020}. There are a number of candidates to explain the softening, given in Refs.~\onlinecite{Park_PRB_2014}. It is likely that the $3a$-CDF we observe is related to the BS phonon anomaly, given both the similarity of the wavevector and its variation with doping. 

\subsection{STM on Bi2212 and the pseudogap}
Spectroscopic imaging scanning tunneling microscopy (SI-STM) measures correlations in the local charge density of states (LDOS).  In Bi$_2$Sr$_2$CaCu$_2$O$_{8+\delta}$ (Bi2212) the LDOS correlations are strongest around $p=0.125$ and weaken as doping is increased, and are no longer detected beyond $x=0.19$~\cite{Fujita_S_2014}. 
The disappearance is accompanied by signatures of the closing of the pseudogap\cite{Fujita_S_2014}. The correlation length of these fluctuations\cite{Hayden2024_HT}, i.e. when converted to $\xi = 1/\sigma$, is around 15\;\AA\ when measured at $T_c$~\cite{Hamidian_NP_2015,Hayden2024_HT}. When measured with RIXS, the correlation length is slightly longer at around 20\;\AA~\cite{Lee_NP_2020,Hayden2024_HT}. The comparison with Bi2212 suggests that the disappearance of the $4a$-CDF in LSCO near $p=p^{\star}$ is associated with the collapse of the pseudogap.

\subsection{Neutron scattering and low-energy spin fluctuations}
The present work suggests the presence of charge density fluctuations (CDF) with relatively low energies $\hbar\Gamma \approx 11$~meV across a large portion of the cuprate phase diagram.   In this section we compare with the collective spin fluctuations.  It is well known that the spin fluctuations in cuprates have several components that evolve with doping and have been extensively characterized.  In the normal state of LSCO, for temperatures $T \approx T_c$ (i.e. comparable with the present measurements), there are strong incommensurate low-energy spin fluctuations (SF) with characteristic energies $\hbar\Gamma \approx 5-10$~meV according to doping \cite{Aeppli_S_1997,Zhu_NP_2022}. The SF are strongly temperature dependent and have been shown to exist into the overdoped region ($x=0.22$) where strange metal behavior occurs and are labeled as ``critical spin fluctuations'' \cite{Radaelli2025_RLZ}.  The wavevectors $\mathbf{Q}_{\text{SF}}$ of the SFs are geometrically related to those of the $4a$-CDF:  $\mathbf{Q}_{\text{SF}}=(1/2 \pm \delta/2, 1/2)$ and $(1/2, 1/2 \pm \delta/2)$.  It is also interesting to note that the SF and $4a$-CDFs have similar correlation lengths with $\xi_{\text{SF}}=20$--25~\AA.  

The fact the the SFs and $4a$-CDFs have have geometrically related wavevectors, similar correlation lengths and similar characteristic energies is consistent with a fluctuating stripe picture\cite{Zaanen1989_ZG, Tranquada2020_Tra}. The stripe picture postulates an entity with intertwined spin and charge modulations. However, it should be noted that the incommensurate spin fluctuations persist to higher dopings $x=0.22$, where the $4a$-CDFs have disappeared at the sensitivity of the RIXS experiments described here. 

\section{Discussion} \label{Sec:discussion}
\subsection{Multi-component nature of the charge correlations}
Calculating the charge density wave order and charge density fluctuations in cuprate superconductors is complex.  The correlations emerge from the multi-orbital Cu-O electronic structure with the oxygen degrees of freedom playing a critical role\cite{Atkinson2015_AKB,Thomson2015_TS,Mai2024_MCMJ}.  We may also need to take into account the coupling to spin fluctuations, hot spots, phonons, material-specific physics, and other factors.  To our knowledge, theory does not yet quantitatively predict the $\delta_{3a}$ and $\delta_{4a}$ CDFs/CDW observed here.  However, multi-orbital calculations do predict\cite{Mai2024_MCMJ} different charge susceptibilities at the Cu and O atoms and $\delta_{\text{charge}} \approx 0.3$. Experimentally, for LSCO there is evidence of charge correlations with $\delta_{4a} \approx 0.24$ from hard x-ray diffraction and RIXS\cite{Croft_PRB_2014,Thampy_PRB_2014,Wen_NC_2019,Miao_nQM_2021,Arx_nQM_2023}.  The $\delta_{3a} \approx 0.3$ correlations are seen in this work at the Cu-$L$ RIXS as a (side-)peak in the intensity and a dip in the energy [Fig.~\ref{fig:phonons}(b)] of the ``BS+$A_{1g}$'' phonon. The frequency of this phonon aligns [see Fig.~\ref{fig:phonons}(a)-(e)] with the BS phonon suggesting that we are primarily seeing this. This is consistent with inelastic neutron scattering and (non-resonant) inelastic x-ray scattering \cite{Park_PRB_2014} studies of the BS phonon in LSCO that show a dip in the phonon energy and a peak in the phonon width (damping) at $h \approx 0.3$. We expect the BS phonon to be particularly sensitive to oxygen charge susceptibility. This suggests the $3a$-CDF is primarily associated with the oxygen site.    

\subsection{Critical fluctuations across the cuprate phase diagram} 
The possibility of a quantum critical point at the critical pseudogap doping $p^{\star}$ has been discussed in many papers\cite{Fujita_S_2014,Michon2019_MGB,Arpaia_NC_2023}. If charge density fluctuations were closely associated with order parameter fluctuations, a diverging correlation length and correlation time would be expected. Our observations do not support this. Instead, we find the correlation length [Fig.~\ref{fig:phase}(d,g,j)] and relaxation rate [Fig.~\ref{fig:energies}(d)] are approximately constant with doping for $x \le 0.19$ with $\xi_{4a-\text{CDF}} \approx 20$~\AA\ and $\hbar\Gamma_{\xi_{4a-\text{CDF}}} \approx 11$~meV. In contrast, the relaxation rate associated with low-energy charge fluctuations decreases rapidly with temperature \cite{Huang_PRX_2021} for $x=0.14$.
Analogous behavior has been observed in the low-energy spin fluctuations in LSCO, where low-frequency ($5-10$~meV) fluctuations are observed at low temperatures ($T \sim T_c$), and these fluctuations appear to be ``critical'' across the phase diagram.  

The extended region of criticality is consistent with a scenario \cite{Radaelli2025_RLZ} in which the cuprate phase diagram is controlled by a quantum Griffiths phase.  A Griffiths phase occurs near a continuous phase transition in systems with quenched disorder. The inclusion of disorder is logical in the La$_{2-x}$Sr$_x$CuO$_4$ system because of the perturbation caused by Sr doping in the plane neighboring the CuO$_2$ plane.  The Griffiths phase corresponds to a smearing of singular behavior as a function of a control parameter. This could naturally lead to slow dynamics and low-frequency relaxation of the collective spin and charge excitations across the phase diagram in LSCO rather than a well-defined quantum critical point.

\subsection{Linear resistivity and Planckian dissipation} 

The present study suggests that the $4a$-CDW and the $4a$-CDFs disappear (or become too weak to observe) for dopings above $x \gtrsim x^{\star} \approx 0.19$. In contrast, the strongly $T$-dependent low-energy spin fluctuations with a geometrically related wavevector ($\delta_{\text{SF}} = \frac{1}{2} \delta_{4a-\text{CDF}}$) are easily observable at higher dopings, e.g. $x=0.22$ \cite{Zhu_NP_2022, Radaelli2025_RLZ}. It is intriguing that in the overdoped region $x \gtrsim 0.19$ where a very clear $T$-linear resistivity (``Planckian dissipation'') \cite{Cooper2009_CWV,Hartnoll2022_HM} only $T$-dependent low-energy spin fluctuations have been observed to date and not the corresponding charge fluctuations. 

\section{Summary} \label{Sec:sum}

In this work we have measured Cu-$L_3$ and O-$K$ RIXS in LSCO over $0.11 < x < 0.22$.  The observed excitations are interpreted using a model in which charge excitations are coupled to phonons.  We find three components ($4a$-CDW, $4a$-CDF, $3a$-CDF) in $(H,E)$-space (see Sec.~\ref{sec:three_components}), each has a distinct a distinct doping evolution. Our data also allows us to comment on the structure of the CDF in $E$. 

The first component is what we have called a $4a$-CDW, which is ordered at $\delta \approx 0.24$, static within energy resolution, and is maximized at $x=0.125$ doping. It has a correlation lengths up to $\xi \sim 50$\;\AA\ and is no longer detected in for $x >0.16$. This is longer-range order that is detected with non-resonant X-rays. The second are the $4a$-CDF, which have the same wavevector as the $4a$-CDW, but a shorter correlation length of around 20\;\AA. The $4a$-CDF couple most strongly to the LA phonon, and we estimate their the energy scale as $\approx 11\;$meV.  The $4a$-CDF are strongest at $x=0.125$, are no longer detected beyond $x=0.19$, and therefore may be associated with the closing of the pseudogap. The energy scale and correlation length of the $4a$-CDF are approximately constant with doping where it can be observed. The third component are the $3a$-CDF, which are also dynamic and have wavevector $\delta \approx 0.29$. The $3a$-CDF exist from $x=0.11$ to $x=0.22$, and show a gradual suppression of intensity with doping. This component can only be seen at the Cu-$L$ edge and may have a higher energy scale of 30--80\;meV because it is seen in a range of phonons. 

Our results show that the charge correlations in LSCO have a more complex structure in $H$-$E$ than previously thought, with ordering tendencies at two different $Q$ values. Each CDW or CDF component shows a distinct evolution with doping.  We find that the $4a$-CDF component disappears at approximately the same doping as the pseudogap. However, there in no evidence of charge fluctuations associated with a quantum critical point.

\section{Acknowledgments}

This work was supported by the UK EPSRC under grant EP/R011141/1. We thank Diamond Light Source for providing beam time under proposal ID MM29150. C.C.T. acknowledges funding from Diamond Light Source and the University of Bristol under joint doctoral studentship STU0372. We thank Seamus Davis for fruitful discussions.

\end{document}